\documentclass[a4paper,11pt]{iopart}
\usepackage{graphicx}
\usepackage{cite}
\usepackage{color}
\usepackage{amsfonts}
\usepackage{srcltx}
\begin{document}

\title{Directed percolation with a single defect site}
\author{A. C. Barato$^1$ and H. Hinrichsen$^2$}
\address{$^1$The Abdus Salam International Centre for Theoretical Physics\\
		 34014 Trieste, Italy\\
         $^2$Universit\"at W\"urzburg,
	 Fakult\"at f\"ur Physik und Astronomie\\
         D-97074 W\"urzburg, Germany}

\ead{acardoso@ictp.it}

\begin{abstract}
In a recent study [arXiv:1011.3254] the contact process with a modified creation rate at a single site was shown to exhibit a non-universal scaling behavior with exponents varying with the creation rate at the special site. In the present work we argue that the survival probability decays according to a stretched exponential rather than a power law, explaining previous observations.
\end{abstract}

%==========================================================================
\section{Introduction}
%==========================================================================

In non-equilibrium statistical physics directed percolation (DP) is known to be the most important universality class of phase transitions into absorbing states, playing a role comparable to the Ising model in equilibrium statistical mechanics~\cite{Kinzel85,MarroDickman99,Hinrichsen00,Odor04,Lubeck04,Odor08a,HenkelEtAl08a}. One of the most intriguing properties of this class is its robustness with respect to the choice of the dynamical rules. Non-DP critical behavior  is obtained only if the dynamical rules are changed substantially, e.g. by introducing additional symmetries, quenched randomness, or algebraically distributed long-range interactions. 

\begin{figure}[t]
\includegraphics[width=160mm]{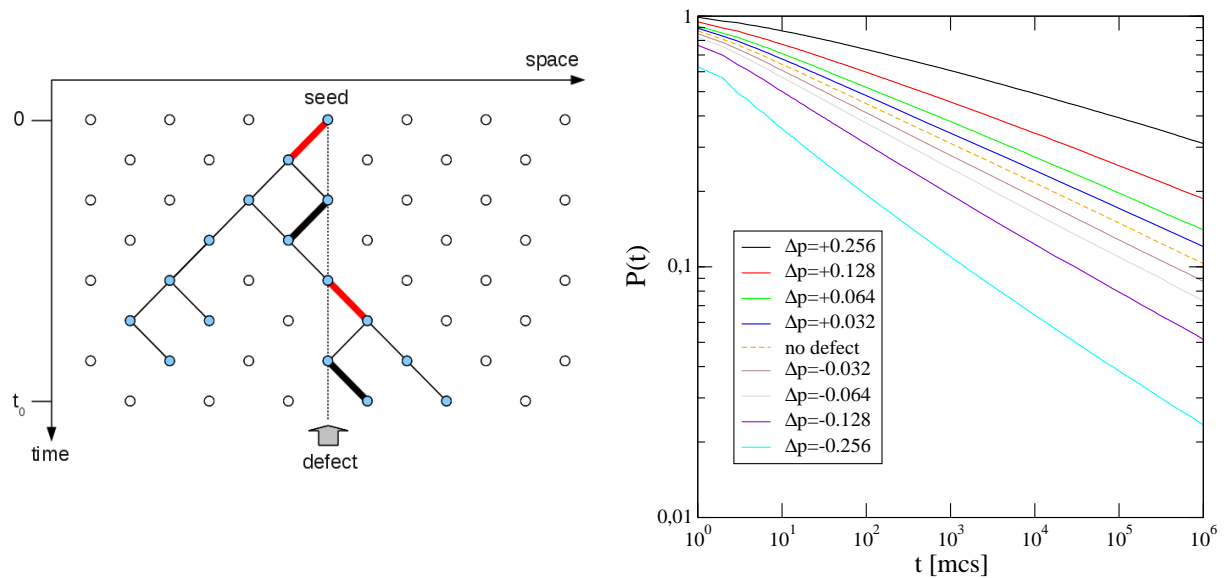}
\caption{Left: Directed bond percolation with a defect. In the bulk each bond is open with probability $p$, while for bonds originating at the defect site (bold line) this probability is modified by $\Delta p$. The figure shows a particular configuration, where a cluster of active sites (blue dots) generated by an active seed at the defect site survives until time $t_0$. Those bonds for which -- if cut -- the cluster would not have survived until $t_0$ are called red bonds and are marked in red color. Right: Decay of the survival probability in a critical bond DP with a defect for various values of $\Delta p$.}
\label{fig1}
\end{figure}

Recently Miller and Shnerb~\cite{MillerShnerb10} studied the one-dimensional contact process with a single defect. The contact process is a version of DP with random-sequential updates following the reaction-diffusion scheme $A\leftrightarrow 2A, A\to \emptyset$ and is often used as a simple model for epidemic spreading. Depending on the rate $\lambda$ for particle creation, it displays a second-order phase transition between a fluctuating phase and the absorbing state (empty lattice) taking place at the critical point $\lambda_c \simeq 3.2978$. Miller and Shnerb implemented a defect at a particular site, where the birth rate $\lambda$ is modified by the amount $\Delta \lambda$. Starting the process with a single active seed at this particular site, they observed that the survival probability $P_{\rm s}(t)$ measured at $\lambda=\lambda_c$ exhibits a power law $P_{\rm s}(t)\sim t^{-\delta}$ with a varying exponent $\delta$ depending on $\Delta\lambda$ (cf. Fig.~\ref{fig1}). This is surprising since the variation of a rate in a \textit{single} site can be considered as a boundary effect which usually does not modify the universality class of the process in the bulk. One should not confuse this with boundary induced phase transitions into an absorbing state \cite{BaratoHinrichsen08}, where the special dynamical rules in the boundary site generates the phase transition.

In this paper we explain these findings. Considering the density of so-called red bonds at the defect site it turns out that the actual decay of the survival probability at criticality does not obey a power law, instead it is dominated by a stretched exponential. Our findings are confirmed by improved numerical simulations.

%==========================================================================
\section{Weak defect approximation}
%==========================================================================

In this section we derive an expression for the survival probability in DP with a weak defect, using the notion of so-called red bonds. Since red bonds are usually discussed in the context of directed bond percolation, we will use this realization of DP instead of the contact process throughout the paper. 

\subsection{Red bonds}
%---------------------

Let us first consider DP without a defect. For a given cluster of sites connected by bonds, a bond is called `red' if its removal would break up the cluster into two pieces. In the context of the present problem, where we study the survival of a DP cluster generated from a single seed, the red bonds of a cluster surviving until time $t$ are those which, if removed, would turn the cluster into a non-surviving one (see Fig.~\ref{fig1}).

In DP, the density of red bonds is closely related to the off-critical behavior of the model. To see this let us consider an ordinary bond-DP process without defects at criticality $p=p_c$ and ask the question how the survival probability will respond to an infinitesimal variation of the global control parameter $p$. Suppose that for a given realization of open and closed bonds a seed at the origin has generated a cluster that survives until time $t$. Let us then slightly decrease the percolation probability by $\Delta p$ and ask for the probability that the cluster turns into a non-surviving one. Clearly, by decreasing $p$ (using the same set of random numbers) some of the conducting bonds of the cluster will be removed. If the removed bond happens to be a red bond, the cluster breaks up into to pieces and becomes non-surviving. Therefore, the response of the survival probability to a variation of $p$ is directly related to the actual number of red bonds.

More specifically, let $N_{\rm tot}(t)$ be the total number of red bonds of the cluster and consider a small change of the percolation probability $p\to p+\Delta p$ by $\Delta p < 0$ in a system without defect. Then the survival probability will respond to lowest order by
\begin{equation}
P_{\rm s}(t,p_c+\Delta p) = P_{\rm s}(t,p_c) (1+\alpha\Delta p)^{N_{\rm tot}(t)} + \ldots
\end{equation}
with some model-dependent constant $\alpha$, so that to leading order the change is given by
\begin{equation}
P_{\rm s}(t,p_c+\Delta p) \sim t^{-\delta} \, \exp(\alpha\,\Delta p\,N_{\rm tot}(t)).
\end{equation}
Since a change of the percolation probability will lead to a breakdown of the algebraic behavior of the survival probability at a time scale of the order $|\Delta p|^{-\nu_\parallel}$, we can immediately conclude that the total number of red bonds in a cluster scales as $N_{\rm tot}(t)\sim t^{+1/\nu_\parallel}$.

\subsection{Red bonds in the center}
%-----------------------------------

Let us now turn to directed percolation with a single defect. Here the situation is analogous to the one described before, the difference being that $p$ is not changed globally but only at the site from where the cluster was generated. Therefore, the response of the survival probability will be determined to leading order by the number of red bonds originating from this particular site, meaning that one simply has to replace the total number of bonds $N_{\rm tot}(t)$ by the number of red bonds in the center, denoted as $N_{\rm c}(t)$. Since we can expect the density of red bonds to be roughly constant in the interior of the cluster, the (time-integrated) number of red bonds in the center will be approximately proportional to $N_{\rm tot}(t)$ divided by the typical spatial volume of the cluster $V\sim t^{d/z}$. This means that $N_{\rm c}(t)$ is expected to scale as
\begin{equation}
N_{\rm c}(t) \sim t^{1/\nu_\parallel - d/z}.
\end{equation}
Using this scaling law we expect the survival probability to behave to leading order as
\begin{equation}
P_{\rm s}(t) \sim t^{-\delta} \, \exp(\alpha\,\Delta p\,t^{1/\nu_\parallel - d/z}) 
\end{equation}
with some non-universal constant $\alpha$. 

\begin{figure}[t]
\centering\includegraphics[width=110mm]{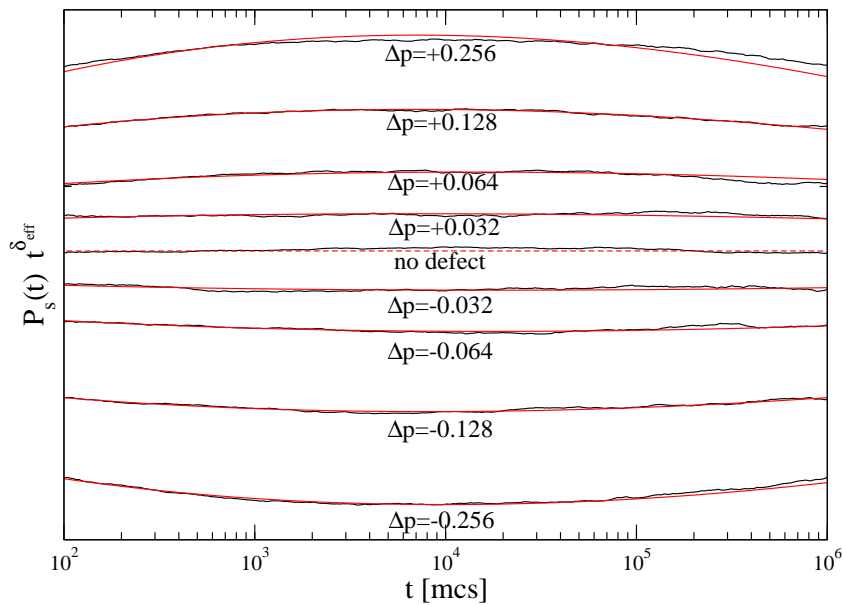}
\caption{Numerical verification of formula~(\ref{eq:fit}). The graph shows the same data as in Fig. 1 multiplied by $t^{\delta_{\rm eff}}$, where $\delta_{\rm eff}$ is the effective survival exponent averaged in the time interval from $10^2$ to $10^6$ time steps. In this representation, clean power laws would appear as horizontal lines. However, the numerical data (black) exhibits a clear curvature depending on the strength of the defect. The suggested stretched exponential~(\ref{eq:fit}) fitted to the data by adjusting the constants $c_1$ and $c_2$ captures this curvature convincingly.
}
\label{fig2}
\end{figure}

%==========================================================================
\section{Numerical results}
%==========================================================================

\subsection{Survival probability}
%-----------------------------------

The argument presented in the previous section implies that the survival probability in 1+1-dimensional DP with a single defect should decay as
\begin{equation}
\label{eq:fit}
P_{\rm s}(t) \simeq c_1 t^{-0.1595} \exp(c_2 \,t^{-0.05586}) \,,
\end{equation}
provided that $\Delta p$ is small and negative. In order to verify this formula, we performed extensive list-based seed simulations in a virtually infinite one-dimensional system, averaging over $10^5$ to $10^6$ runs depending on $\Delta p$ and over $10^6$ time steps. Note that the simulations in Ref.~\cite{MillerShnerb10} cover only 30.000 time steps, - a much smaller range where the curvature of the data in loglog representation was probably not easy to see.

To demonstrate the non-algebraic behavior, we first determined the effective exponent $\delta_{\rm eff}$ by an ordinary power-law fit over four decades of time ranging from $10^2$ to $10^6$ Monte Carlo steps (see Table 1). Then we multiplied the data by $t^{\delta_{\rm eff}}$: In a loglog representation a clean power law would give a horizontal line so that even tiny deviations are much easier to see. In fact, as shown in Fig.~\ref{fig2}, there is a systematic curvature depending on the strength of the defect. 

A non-linear fit of Eq.~(\ref{eq:fit}) over four decades reproduces this curvature accurately. It is surprising that the formula, which was derived in the limit of small and negative $\Delta p$, works very well for positive $\Delta p$ and  even up to $|\Delta p|= 0.256$, which is more than 30\% away from the critical point. The corresponding estimates of the constants $c_1$ and $c_2$ are listed in Table~1.

\begin{table}
\begin{center}
\begin{footnotesize}\begin{tabular}{|c||c|c|c|c|}
\hline
$\Delta p$ & $\delta_{\rm eff}$ & $c_1$ & $c_2$ & $c_2/\Delta p$\\ \hline
$-0.256$ & 0.230 & 0.078(5) & 2.12(5)  &  -8.3 \\
$-0.128$ & 0.195 & 0.29(2)  & 1.05(3)  &  -8.2 \\
$-0.064$ & 0.177 & 0.52(1)  & 0.53(2)  &  -8.3 \\
$-0.032$ & 0.168 & 0.70(1)  & 0.26(2)  &  -8.1 \\
$0$	 & 0.159 & 0.95(1)  & 0.01(2)  &  --   \\
$+0.032$ & 0.150 & 1.24(2)  & -0.27(2) &  -8.4 \\
$+0.064$ & 0.142 & 1.63(3)  & -0.53(3) &  -8.3 \\
$+0.128$ & 0.126 & 2.68(5)  & -1.01(4) &  -7.9 \\
$+0.256$ & 0.094 & 6.8(1)   & -1.94(5) &  -7.6 \\
\hline
\end{tabular}
\end{footnotesize}
\caption{Numerical estimates of the effective exponent of the survival probability (average slope of the data without error margin) and the fit parameters $c_1$ and $c_2$ for various values of the defect strength $\Delta p$ (see text).}
\label{tab}
\end{center}
\end{table}

As an additional check, we verify the prediction that the second fit parameter $c_2$ should be proportional to $\Delta p$. In fact, the ratio $c_2/\Delta p$ listed in Table 1 is approximately constant except for large values of $\Delta p$, where we see a small deviation.

\subsection{Number of active sites}
%-----------------------------------

Besides the survival probability, seed simulations provide data for the total number of active sites $N(t)$ averaged over all runs and the mean square spreading from the origin $R^2(t)$ averaged over all surviving tuns. In critical DP without defects, these quantities scale as $N(t)\sim t^\eta$ and $R^2(t)\sim t^{2/z}$, where $\eta=1/z-2\delta$ and $z=\nu_\parallel/\nu_\perp$.	

\begin{figure}[t]
\centering\includegraphics[width=120mm]{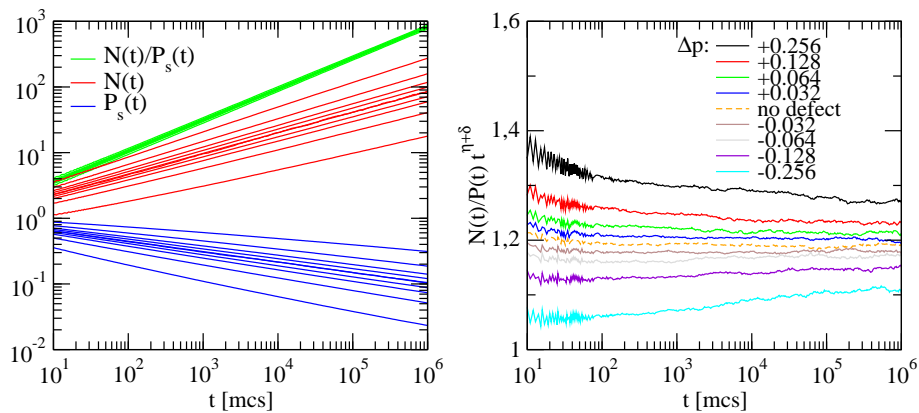}
\caption{The number of active sites $N(t)$, the survival probability $P_s(t)$, and the ratio $N(t)/P_s(t)$ in seed simulations of DP with a single defect at the origin.
}
\label{fig3}
\end{figure}

As shown in Fig.~\ref{fig3}, the quantity $N(t)$ shows similar deviations like $P_s(t)$ if the defect strength $\Delta p$ is varied. However, the quotient $N(t)/P_s(t)$, which can be interpreted as the number of active sites averaged over surviving runs only, seems to exhibit clean power laws. This quotient divided by the expected power law $t^{\eta+\delta}$ is shown on the right hand side of the figure. As can be seen, small deviations of next leading order are still visible. Nevertheless the quotient $N(t)/P_s(t)$ approximates a power law much better than $N(t)$ and $P_s(t)$ in itself.

\subsection{Density at the defect site}
%--------------------------------------

DP with a defect can also studied by simulations with homogeneous initial states. For example, we may ask how the density at the defect site decays as a function of time starting with a fully occupied lattice. It turns out that this density decays in the same way as the survival probability analyzed above. This is no surprise since rapidity reversal, a symmetry implying $P_s(t)$ and the density at the "seed site" starting with fully occupied lattice to decay with the same exponent, holds also in presence of a defect. A proof of this relationship is given in Appendix A.

%==========================================================================
\section{Summary}
%==========================================================================

In this work we have shown that a single defect in 1+1-dimensional DP does not change the universality class of the process, opposed to the conclusion obtained in Ref.~\cite{MillerShnerb10}. For weak defects with $\Delta p/p\ll 1$ we derived a formula for the survival probability by considering the density of red bonds at the defect site. 

We note that in the extreme case $\Delta p=-p$ the defect site acts as an absorbing boundary. Even in this case the universality class of the bulk process does not change, instead one obtains an interesting boundary critical behavior with a new surface critical exponent~\cite{FroedhEtAl98a,FroedhEtAl01a}. Away from this point, this boundary critical phenomenon breaks down and the defect site has no influence in the renormalization group sense, instead it manifests itself as a correction of certain quantities such as the survival probability in form of a stretched exponential.

%==========================================================================
\appendix
\section{Time-reversal symmetry}
%==========================================================================

It is well known that directed bond percolation without defects has the special property that the time reversal symmetry is exact. It can be checked easily that this proof can be generalized to percolation probabilities with any type of quenched disorder, including the single defect as a special case. 

In order to make sure that this symmetry is not an artifact of directed bond percolation but valid in any realization of DP, we show here that the time-reversal symmetry also holds in the contact process for transition rates $\lambda$ depending on the lattice position. The calculation below is similar to the proof of time-reversal symmetry for the homogeneous contact-process presented in~\cite{Grassberger79}.   

As in Ref.~\cite{Grassberger79} we define the creation and annihilation operators 
\begin{equation}
 \hat{a}_i|s_i\rangle= s_i|s_i-1\rangle\qquad,\qquad \hat{a}_i^\dagger|s_i\rangle= (1-s_i)|s_i+1\rangle,
\end{equation} 
where $s_i= 0,1$. The operator $\hat{H}$ for the CP is given by 
\begin{equation}
\hat{H}= \sum_i[(\hat{a}_i^\dagger-1)\hat{a}_i]+\sum_{<ij>}\lambda_i[(\hat{a}_i-1)\hat{a}_i^\dagger\hat{a}_j^\dagger\hat{a}_j]. 
\end{equation}
The density of particles at some site $i$ is given by 
\begin{equation}
\rho_i(t)= \langle0|\hat{a}_i\hat{V}|\psi(t)\rangle
\end{equation} 
where $\hat{V}= \prod_j(1+\hat{a}_j)$. Note that only configurations with a particle at $i$ "survive" the action of the operator $\hat{a}_i\hat{V}$. If the initial condition is a fully occupied lattice we have 
\begin{equation}
\rho_i(t)= \langle0|\hat{a}_i\hat{V}\exp(-\hat{H}t)\prod_j{\hat{a}^\dagger_j}|0\rangle.
\label{c3thisequation}
\end{equation} 
The pseudo-Hermitian operator $\bar{H}= \hat{V}\hat{H}\hat{V}^{-1}$ obeys the relation \cite{Grassberger79}
\begin{equation}
\bar{H}^\dagger= \hat{\sigma}\bar{H}\hat{\sigma},
\end{equation} 
 where $\hat{\sigma}= \hat{\sigma}^\dagger= \hat{\sigma}^{-1}= \prod_j[\hat{a}_j,\hat{a}^\dagger_j]$ and $\hat{V}^{-1}= \prod_j(1-\hat{a}_j)$. Note that a possible inhomogeneity of the creation rate has no influence on the last equality.   

Finally, Eq.~(\ref{c3thisequation}) implies that
 \begin{equation}
\rho_i(t)=  \langle0|\hat{a}_i\exp(-\bar{H}t)\hat{V}^\dagger|0\rangle=\langle0|V\hat{\sigma}\exp(-\bar{H}t)\hat{\sigma}\hat{a}_i^\dagger|0\rangle. 
 \end{equation} 
 Since $\hat{V}^\dagger\hat{\sigma}\hat{V}^\dagger|0\rangle= |0\rangle$ and $\hat{\sigma}\hat{a}_i^\dagger|0\rangle= -\hat{a}_i^\dagger|0\rangle$, we obtain the final result
 \begin{equation}
\rho_i(t)= -\langle0|\exp(-\hat{H}t)\hat{V}^{-1}\hat{a}_i^\dagger|0\rangle= 1-\langle0|\exp(-\hat{H}t)\hat{a}_i^\dagger|0\rangle= P_s(t).
\end{equation}
This proves the time reversal symmetry in the contact process with position-dependent creation rates.

The fact that time reversal symmetry holds provides a direct way of observing that a single defect site does not alter the universality class. If the density of active sites (averaged over the whole lattice) starting with a fully occupied lattice is measured we expect a clean DP decay exponent, which imply in the single defect site generating a transient effect that is felt near enough the boundary.

%==========================================================================
\section*{References}
%==========================================================================

% \bibliographystyle{iopart-num}
% \bibliography{/home/hinrichsen/Dokumente/Literatur/master}

\providecommand{\newblock}{}

\end{document}